\documentclass{amsart}%
\usepackage{amsfonts}
\usepackage{amsmath}
\usepackage{amssymb}
\usepackage{graphicx}%
\providecommand{\U}[1]{\protect\rule{.1in}{.1in}}

\theoremstyle{plain}

\numberwithin{equation}{section}
\begin{document}
\title[Symmetry-Breaking]{Symmetry-Breaking in Plant Stems}
\author{Larry Harper}
\address[A. One]{Department of Mathematics\\
University of California, Riverside}
\email{harper@math.ucr.edu}
\author{Greg Huey}
\address[A. Two]{Department of Mathematics\\
University of California, Irvine}
\email{hueyg@uci.edu}
\date{October 27, 2021}
\dedicatory{Muchas gracias a Juan y Celia Gonzalez de Pereira, Colombia. Desde el
balc\'{o}n de la villa Gonz\'{a}lez Me contempl\'{o} un exuberante mar de los
tallos de las plantas. Me comenz\'{o} preguntar una vez m\'{a}s sobre los
procesos en los que se basan sus innumerables variaciones exquisitas.}
\begin{abstract}
The purpose of this paper is to present a model of a phenomenon of plant stem
morphogenesis observed by Cesar Gomez-Campo in 1970. We consider a simplified
model of auxin dynamics in plant stems, showing that, after creation of the
original primordium, it can represent random, distichous and spiral
phyllotaxis (leaf arrangement) just by varying one parameter, the rate of
diffusion. The same analysis extends to the $n$-jugate case where $n$
primordia are initiated at each plastochrone. Having validated the model, we
consider how it can give rise to the Gomez-Campo phenomenon, showing how a
stem with spiral phyllotaxis can produce branches of the same or opposite
chirality. And finally, how the relationship can change from discordant to
concordant over the course of a growing season.

\end{abstract}
\maketitle

\section{ Introduction}

\subsection{"As the Twig is Bent,..."}

As a student (more than fifty years ago) the senior author came across a
popular exposition on the mathematical patterns in phyllotaxis (leaf
arrangement) and was fascinated. The Fibonacci numbers, $F_{n}$, are a
mathematical sequence that begins with $F_{1}=1$, $F_{2}=2$ and then for
$n\geq2$, $F_{n+1}=F_{n}+F_{n-1}$. So the numbers continue%
\[%
\begin{tabular}
[c]{r||rrrrrrrrrrr}%
$n$ & $1$ & $2$ & $3$ & $4$ & $5$ & $6$ & $7$ & $8$ & $9$ & $10$ &
$11...$\\\hline
$F_{n}$ & $1$ & $2$ & $3$ & $5$ & $8$ & $13$ & $21$ & $34$ & $55$ & $89$ &
$144...$%
\end{tabular}
\]
\textit{ad infinitum}. That Fibonacci numbers actually occur in phyllotactic
structures on pineapples and pine cones and many other plants inspired awe and
curiosity about what lay behind it. It seemed that the mechanism must be
stable but flexible and therefore probably simple. At that time (1960) so
little was known about morphogenesis that insight into the underlying
processes seemed pure fantasy. However, things have changed.

\subsection{Background}

\subsubsection{Hofmeister's Rule}

Much has been written about phyllotaxis since Bravais and Bravais initiated
its mathematical study in 1837. Being amateurs in botany we have taken the
books by Roger V. Jeans (\cite{Jea94}, \cite{Jea98}) as authoritative sources
for the pre-21$^{st}$ century literature of phyllotaxis. The major insight of
the 21$^{st}$ century has been the central role of the plant hormone, auxin.
Since primordia (points on a stem where new organs are being initiated) tend
to distance themselves from each other. This is known as Hofmeister's Rule
(See the website%
\[
\text{http://www.math.smith.edu/phyllo/About/math.html}%
\]
or \cite{R-P}, p.102). It was long thought that primordia must be the source
of a hormone inhibiting other primordia. Experiments have shown however, that
the hormone auxin, which stimulates cells to grow and reproduce (mitosis), is
responsible for Hofmeister's Rule. A new primordium becomes an auxin sink
because it utilizes auxin to build the vascular structure underneath to
support its further growth. Auxin is created at various rates in all plant
tissues so it is the depletion of auxin near a primordium that inhibits the
establishment of new primordia nearby. Since a source of an inhibitory hormone
and a sink of a stimulating hormone have essentially the same effect and the
same dynamics, we have found it easier to maintain the original model,
thinking of a primordium as a source of antiauxin.

\subsubsection{The Model of Smith, \textit{et al}}

To be effective, theory requires a dialog with experiment. The field theory of
auxin dynamics has been vetted by various laboratory experiments: Surgery and
application of hormones known to promote or counter the effects of auxin. In
2004 Smith, \textit{et al} proposed a mathematical model of a plant stem
\cite{SGMRKP} that we call the SGMRKP model.%
\begin{figure}
\centering
\includegraphics[keepaspectratio, width=5.00in]{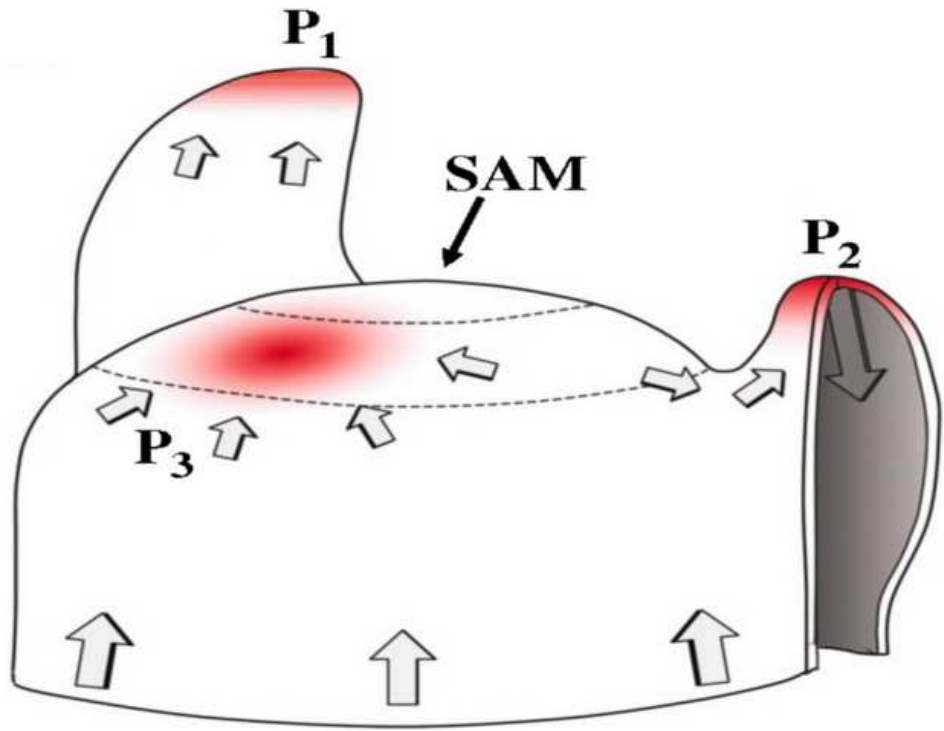}
\caption{Model of a developing plant stem. P$_{1},$P$_{2},$P$_{3}$ are primordia.}
\end{figure}

Most of the morphogenetic activity in a plant stem takes place near the tip,
called the \textit{meristem} and in the outer layer of cells, called the
\textit{tunica}. The tunica is a single layer of cells overlaying the vascular
and supporting structure underneath called the \textit{phloem}. Auxin is
created and consumed at various rates in various tissues. At the apex of the
tunica is a group of (from 1 to 40) pluripotent (stem) cells called the
\textit{stem apical meristem }(SAM). New organs (\textit{primordia}) such as
leaves or flowers are initiated at points where auxin concentration reaches a
critical value. Primordia then absorb auxin from the surrounding tissues in
order to build the underlying vascular structure for continued growth. Thus a
primordium becomes an auxin sink which inhibits the initiation of other
primordia in its immediate vicinity. The SAM itself is a site of growth,
extending the stem, and therefore an auxin sink. This means that primordia can
only be initiated on its periphery (the \textit{peripheral zone}). Auxin is
transported by diffusion but also actively by cells that are polarized to pump
it in one side and out the other. Smith \textit{et al} proposed a set of
equations for auxin dynamics (the SGMRKP model). They were able to solve the
equations numerically (on the computer) and create movies of stems growing in
various standard phyllotactic patterns - see the website
https://www.jic.ac.uk/research-impact/genes-in-the-environment/
 of Smith's group 
Genes in the Environment
at The John Innes Centre.
Figure 2 is a photo of a real plant
stem, copied from
\[
\text{http://www.cobalt-group.com/frontpagewebs/Content/Classify/}%
\]%
\begin{figure}
\centering
\includegraphics[keepaspectratio, width=5.00in]{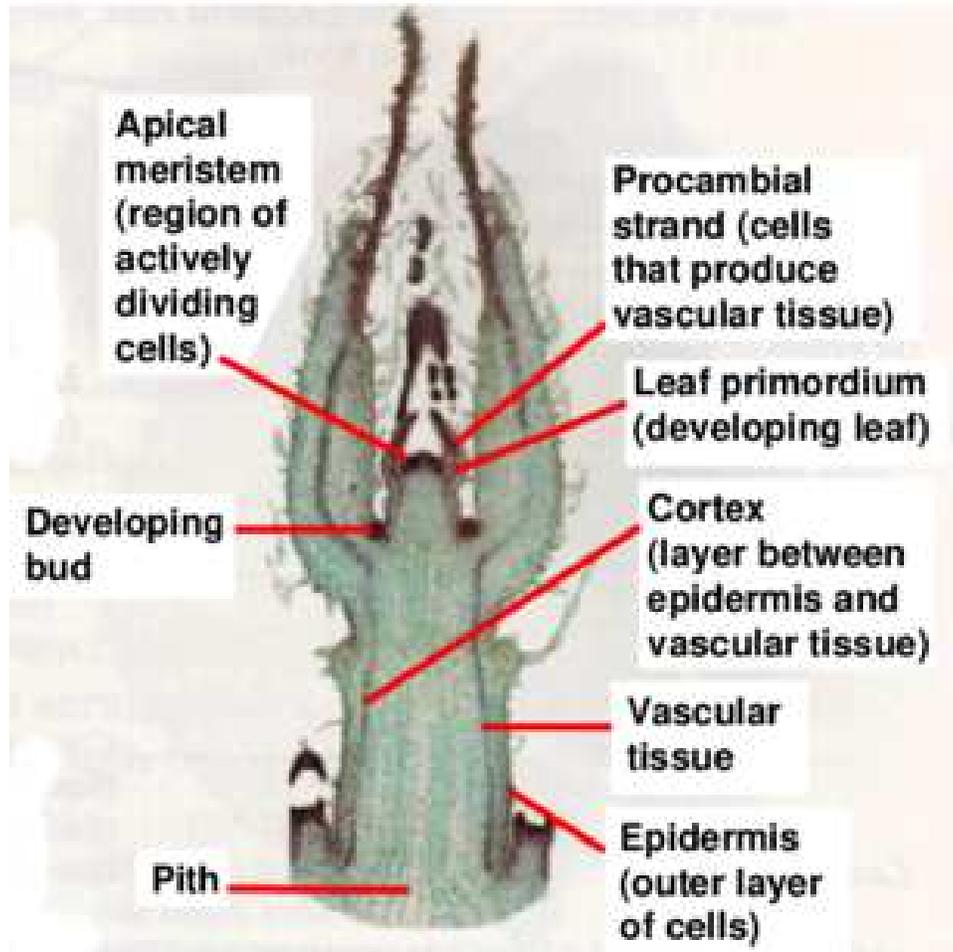}
\caption{A real plant stem}
\end{figure}
The SGMRKP model of a growing plant stem leaves out many characteristics of
real stems. The hope is that it does capture some vital essence.

\subsubsection{Davis's Hypothesis}

The first inkling that the present authors had of being able to make progress
in understanding the processes underlying phyllotaxis came \textit{circa} 1980
from a lecture given in the UCR Mathematics Colloquium by T. Anthony Davis.
Dr. Davis was a coconut palm inspector in Kerala Province, India. Inspecting
coconut palms (\textit{cocos nucifera}) was routine work, so Davis began
looking into phyllotaxis and noted that, like all plants with spiral
phyllotaxis, coconut palms come in two types, left and right handed.
Basically, new leaves are formed one at a time in the crown of the tree, each
leaf rotated approximately 137.5$^{\circ}$ from the previous one. If that
upward spiral goes to the right, the palm is called right handed. If to the
left, left handed. That original arrangement, if left handed, is $(1,2)$
phyllotaxis, the $1$ (genetic) spiral to the left and $2$ spirals of
contiguous primordia,called contact parastichies, to the right. If right
handed we denote it as $\left(  2,1\right)  $ phyllotaxis. As a stem grows,
the primordia move around to accomodate alterations in proportion and change
contiguous neighbors in a systematic way. As a result, the numbers of
(left,right) spiraling parastichies changes from ($F_{n-1},F_{n}$) to
($F_{n+1},F_{n}$) (or ($F_{n},F_{n-1}$) to ($F_{n},F_{n+1}$)). Pine cones and
pineapples often have ($8,13$) or ($13,8$) phyllotaxy. The genetic spiral will
no longer be a contact parastichy, but its direction will be the same as that
of the odd order contact parastichies whose numbers are $1,3,8,21$, $55$,
\textit{etc. }A few observations of almost any variety of plant with spiral
phyllotaxis will show that both chiralities are possible. Are they then
equally probable? What effect does the handedness (chirality) of a palm have
on the health \& productivity of the tree? Those were the questions with which
Dr. Davis entertained himself and claimed to answer: He concluded that they
are not equally probable. Of the over 5,000 coconut palms Davis had examined,
51.4\% were left handed. By statistics, this was an unlikely event (if the
underlying probabilities were equal, 1/2 and 1/2), so the hypothesis (of
equality) should be rejected. He went on to gather data from coconut palm
inspectors around the world and concluded that those coconut palms in the
northern hemisphere were predominantly left handed and those in the southern
hemisphere predominantly right handed \cite{Dav}.

\section{First Insights}

\subsection{Vetting Davis's Hypothesis}

Davis's work was intriguing because if enviromental factors, such as the
hemisphere a specimen happens to grow in, could effect chirality, then it
opened a window through which morphogenetic processes might be illuminated.
And testing of the theory could be done by amateurs! No labs or expensive
machinery would be needed. Southern California has lots of palm trees, not
coconut palms but why should that make a difference? Also, Riverside is quite
a bit further north ($33^{\circ}$) than Kerala Province ($5^{\circ}$ to
$15^{\circ\text{)}}$ and that should make the effect even greater (and
therefore easier to detect). Furthermore, the discovery of DNA by Crick and
Watson in 1953 had set the stage for one of the biggest questions in science
today: How does the abstract (digitally encoded) information on DNA get
translated into the form and function of a living organism? Though probably a
long and complex process, its unraveling will certainly go more quickly if we
can pull it apart from both ends. Davis's hypothesis opened up the possibility
of working backward from the end product of development to make inferences
about the underlying physico-chemical processes.

The most common local palms, iconic for the Los Angeles area, are
\textit{Washingtonia Robusta. }They are tall (50 to 100 feet) with smooth
trunks. This makes it difficult to determine the chirality of their
phyllotaxis (a problem that Davis also had to deal with). However, We soon
found alternatives in \textit{Xylosma Senticosa} and bottlebrush (genus
\textit{Callistemon)}. Both are plentiful on the UCRiverside campus, their
stems have spiral phyllotaxis and chirality is relatively easy to determine.
We spent days collecting data and analysing it. We mentored undergrad research
projects doing the same. However, the results were not satisfying. Our data
consistently indicateded that left handers do predominate but only by a slim
margin near Davis's 51.4\%. Davis's hypothesis implied that the difference
should be greater at our lattitude. We thought, "Maybe the Davis effect
requires vertical stems" (Xylosma and bottlebrush stems grow in all
directions). So we collected data from lettuce and \textit{Washingtonia}
seedlings (both in a nursery setting). Again, the results were similar except
that the more data we collected and the more proficient we became at
collecting it, the smaller the predominance of left handers became. Was this
an indication of observer bias? Evidently shrinking rejection statistics are
common in those areas of science where statistical methods are applied.
Professor Robert Rosenthal of the UCR Psychology Department, an expert on this
phenomenon, has counciled caution against rejecting the rejection of null
hypotheses when shrinking rejection statistics are encountered. Ultimately
however, we decided that Davis's hypothesis was not suitable for our purposes:
Even if the chirality of a stem is effected by lattitude, the effect is small
and not convenient for amateur experimentation.

\subsection{The Gomez-Campo Phenomenon}

However, in the course of vetting Davis's hypothesis, we came upon a way to
influence the chirality of a stem that could be useful: In some plants with
spiral phyllotaxis (such as bottlebrush, but not palms) the primary stem gives
rise to secondary stems (branches). Under certain circumstances a leaf will
produce a new stem (called an axillary bud) from the upper side of its base.
Secondary stems are functionally the same as primary stems. Does a branch have
the same chirality as its parent stem? Again, a few observations on
bottlebrush showed that the answer is "No". Is the chirality (like that of a
primary stem) random (50/50)? A few more observations with bottlebrush showed
that the chirality of a branch is the same as that of its parent stem about
2/3 of the time. Following up on this in the library (circa 1982) we found
that the phenomenon was already known. A Spanish botanist, Cesar Gomez-Campo,
had studied it in 6 annual shrubs in the environs of Madrid and found that the
correlation between the chirality of a stem and a branch could even change
over the course of the growing season: It could be positive at one time (as it
was overall with our bottlebrush) and negative at another \cite{G-C}.

So what mechanism could underly Gomez-Campo's observations? As before, it had
to be stable, yet flexible (and so probably fairly simple). But now it had to
be partly deterministic and partly stochastic. This really intrigued us, but
again we had no idea as to how to answer the question. Over the years a
possible answer slowly took shape: Relative rates of diffusion of auxin
interacting with the geometry of the stem. But before we can apply that
insight to the Gomez-Campo phenomenon, we need to examine our conceptual model
more closely, to clear up an apparent paradox.

\subsection{The Paradox of Distichous Phyllotaxis}

Hofmeister's Rule has been the basis for much of the theory of Fibonacci
phyllotaxis. In the case where primordia appear one at a time (unijugate
phyllotaxy), it implies that the next primordium will appear in the available
space (the peripheral zone) at a point farthest from previous ones. This works
nicely for spiral phyllotaxis, but what about the distichous phyllotaxis of
Clavia (Figure 3), grasses and bird-of-paradise?
\begin{figure}
\centering
\includegraphics[keepaspectratio, width=5.00in]{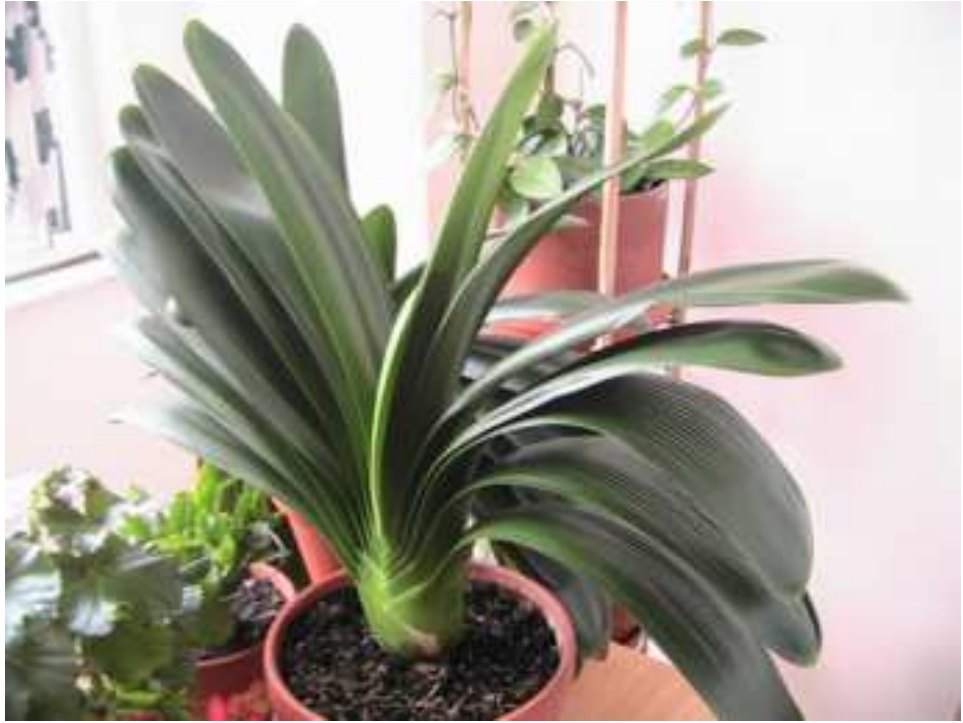}
\caption{The distichous phyllotaxis of Clavia}
\end{figure}
In distichous phyllotaxis, two successive leaves on one side appear to be
closer to each other than the intermediate one on the other side, apparently
contradicting Hofmeister's Rule. Hofmeister himself "explained" this by noting
that the bases of distichous leaves wrap around the stem to the other side,
shielding the following primordium from influence of the leaf preceding it.
However we do not find this explanation to be entirely satisfying. Primordia
are usually initiated at one point. Why does the base of a distichous leaf
extend so much further than in other plants?

\subsection{A Quantitative Model}

Think of the plant stem as a cylinder of semicircumference 1. A primordium
will beome an antiauxin source and we shall assume only diffusion for
transport. Let the unit of time (the plastochrone) be the time to the creation
of the next primordium and let $\sigma^{2}$ be the variance of the density of
antiauxin in the peripheral zone at that time. The normal (Gaussian) density
function at time $t$ for diffusion on the real line starting from a unit
source at $x_{0}$ is
\[
f\left(  x;x_{0},\sigma,t\right)  =\frac{1}{\sqrt{2\pi t}\sigma}%
e^{-\frac{\left(  x-x_{0}\right)  ^{2}}{2\sigma^{2}t}}.
\]
\ For $x_{0}=0$, $\sigma=1$ and $t=1$ the graph of this density is the
familiar bell-shaped curve (Figure 4).%
\begin{figure}
\centering
\includegraphics[keepaspectratio, width=5.00in]{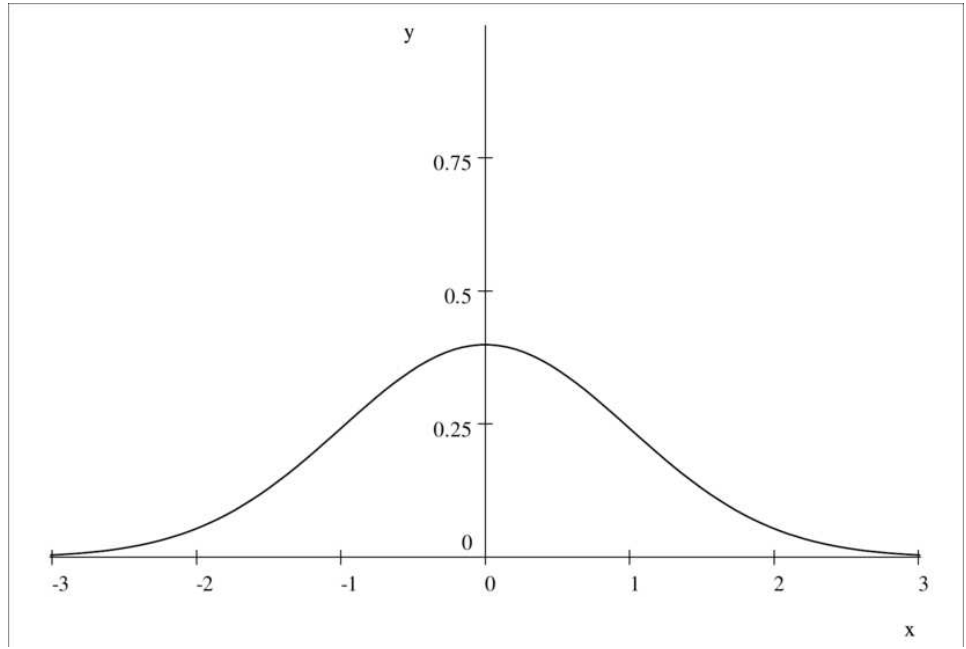}
\caption{Graph of $f\left(  x;0,1,1\right)  =\frac{1}{\sqrt{2\pi}} e^{-\frac{x^{2}}{2}}$, $-3<x\leq3$}
\end{figure}

The values of $x$ extend mathematically to $\pm\infty$, but according to the
table of values for the complimentary error function, erfc (See Wikipedia),
the area under the curve for $x>3$ and $x<-3$ is erfc$(3)=2.21\times10^{-5}$,
a negligeable amount.

Diffusion is a linear process, so if we take this same diffusion process and
put it on the circle of semicircumference $1$ (the peripheral zone), the tails
of the distribution will wrap around and add to give a density of
\[
f\left(  x;0,1,1\right)  +f\left(  x;2,1,1\right)  +f\left(  x;-2,1,1\right)
,-1<x\leq1.
\]
Higher order wraparounds are negligeable and the graph is displayed in Figure
5.%
\begin{figure}
\centering
\includegraphics[keepaspectratio, width=5.00in]{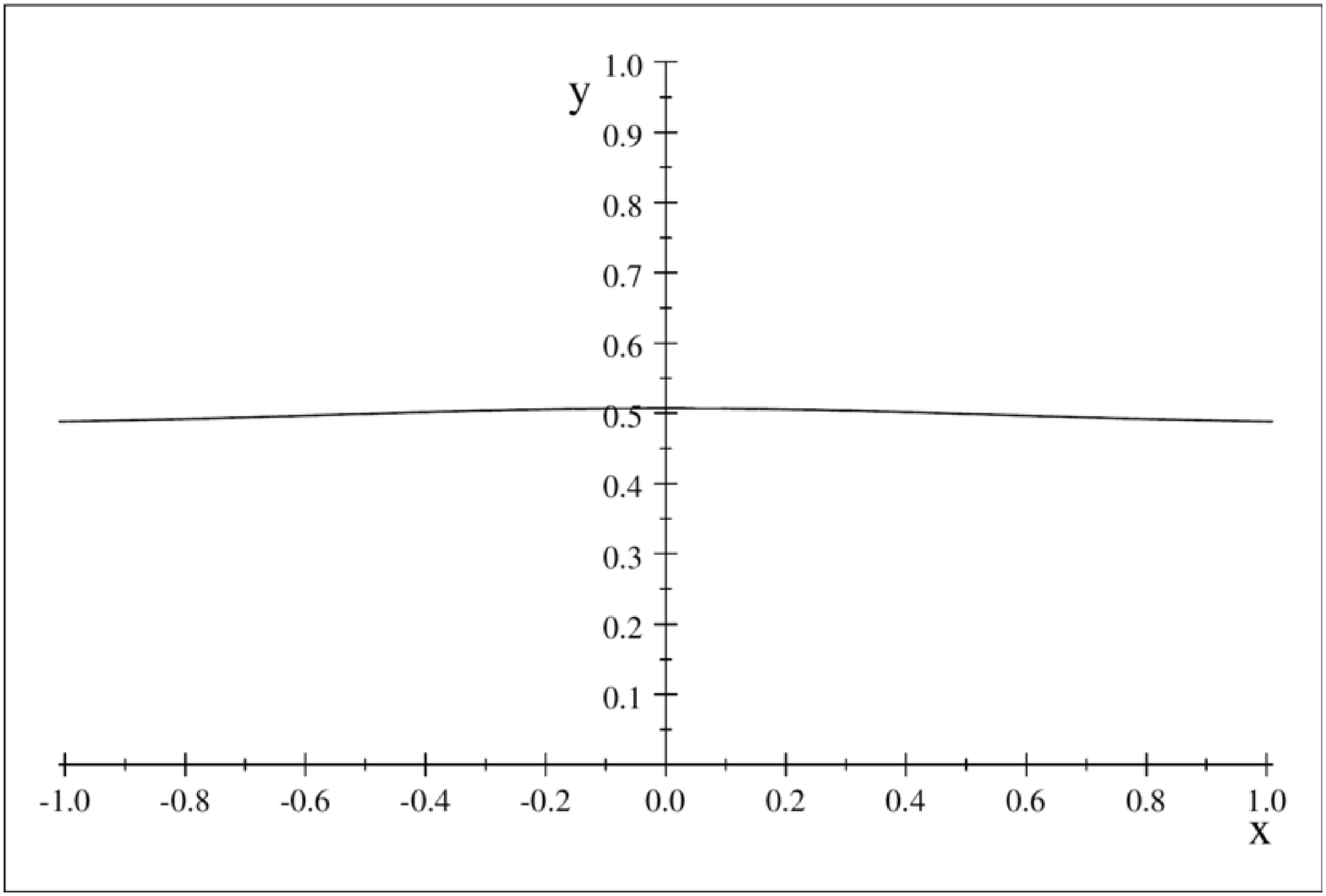}
\caption{Graph of $f\left(  x;0,1,1\right)  +f\left(  x;2,1,1\right) +f\left(  x;-2,1,1\right)  ,$ $-1<x\leq1$}
\end{figure}
Somewhat unexpected (but not inconsequential) is that this function is almost
constant (near .5). We assume that the placement of the first primordium on a
stem is arbitrary. Small random fluctuations in the (theoretically constant)
density of antiauxin will determine where it goes. Now if $\sigma=1$ and a
second primordium is established one plastochrone after the first, its
placement will also be essentially random because the density of antiauxin is
again constant, and so on. This accounts for random phyllotaxy (Again, see
\cite{R-P} p. 102), a phenomenon that we had previously ignored because it
seemed to "lack structure" and contradict Hofmeister's Rule.

\subsubsection{Distichous Phyllotaxy}

If we start with a density that has $\sigma=1/2$ on the same circle (of
semicircumference $1$) we get a density function,
\[
\frac{2}{\sqrt{2\pi}}\left(  e^{-2x^{2}}+e^{-2\left(  x-2\right)  ^{2}%
}+e^{-2\left(  x+2\right)  ^{2}}\right)  ,-1\leq x\leq1,
\]
whose graph is Figure 6
\begin{figure}
\centering
\includegraphics[keepaspectratio, width=5.00in]{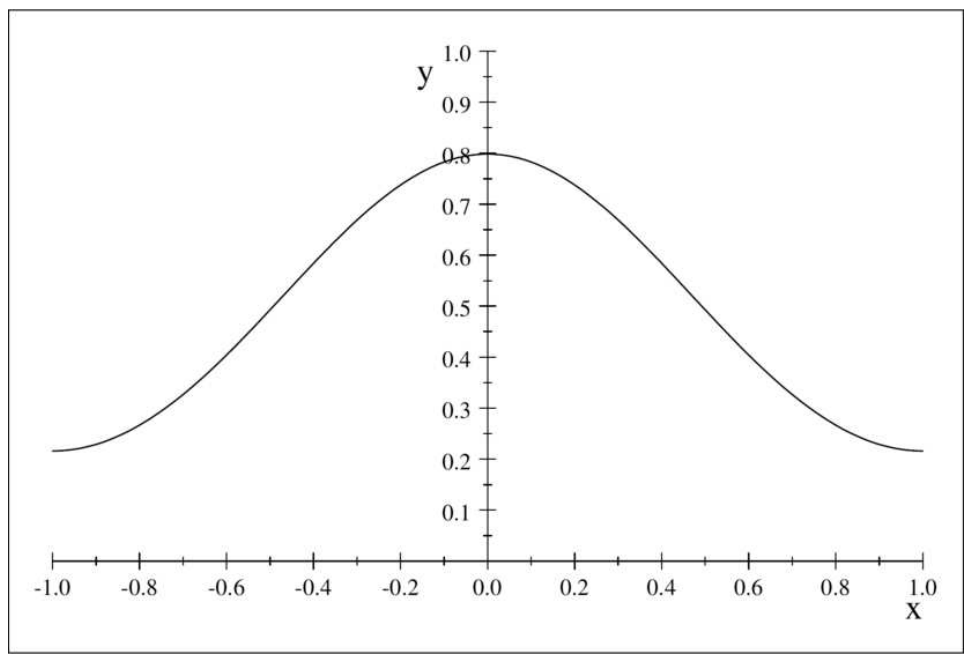}
\caption{Density on the circle with $\sigma=1/2$}
\end{figure}

Unlike the case $\sigma=1$, this distribution (of antiauxin) has a definite
minimum at $x=\pm1$, the antipode of $x=0$. After one unit of time (the
plastichrone) a 2$^{nd}$ primordium is created at $\pm1$. If we recenter
Figure 6, translating in $x$ to bring the minimum back to the origin, then
after another unit of time the density will be
\begin{align*}
&  \frac{2}{\sqrt{2\pi}}\left(  e^{-2x^{2}}+e^{-2\left(  x-2\right)  ^{2}%
}+e^{-2\left(  x+2\right)  ^{2}}\right) \\
&  +\frac{1}{\sqrt{\pi}}\left(  e^{-\left(  x+1\right)  ^{2}}+e^{-\left(
x-1\right)  ^{2}}+e^{-\left(  x+3\right)  ^{2}}+e^{-\left(  x-3\right)  ^{2}%
}\right)  ,
\end{align*}
whose graph is Figure 7.
\begin{figure}
\centering
\includegraphics[keepaspectratio, width=5.00in]{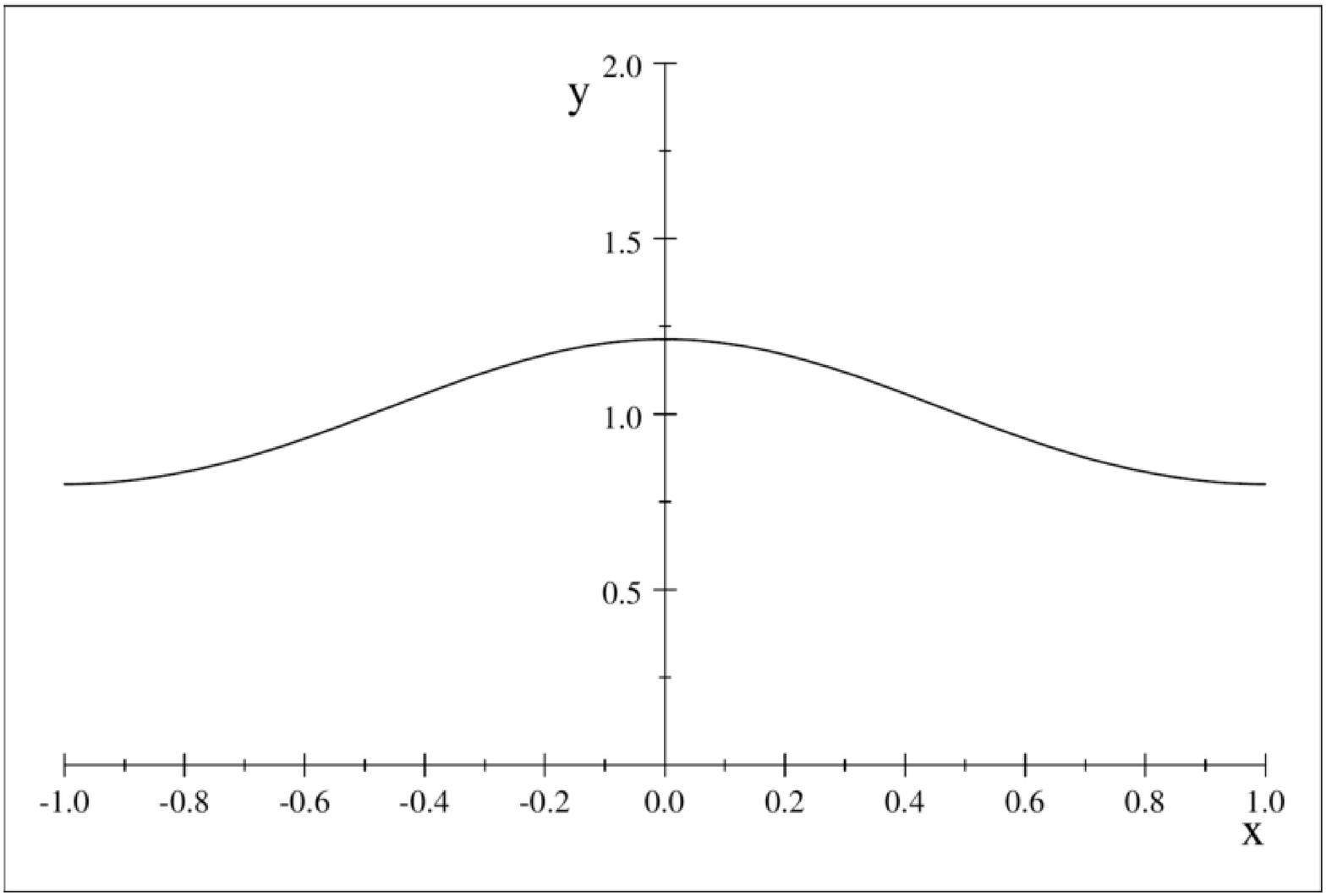}
\caption{$\sigma=1/2$ density after 2$^{nd}$ primordium}
\end{figure}

So there is still a definite minimum at $\pm1$. If we add a $3^{rd}$
primordium at that minimum and rotate again, after another unit of time the
distribution will be%

\begin{align*}
&  \frac{2}{\sqrt{2\pi}}\left(  e^{-2x^{2}}+e^{-2\left(  x-2\right)  ^{2}%
}+e^{-2\left(  x+2\right)  ^{2}}\right) \\
&  +\frac{1}{\sqrt{\pi}}\left(  e^{-\left(  x+1\right)  ^{2}}+e^{-\left(
x-1\right)  ^{2}}+e^{-\left(  x+3\right)  ^{2}}+e^{-\left(  x-3\right)  ^{2}%
}\right) \\
&  +\frac{2}{\sqrt{6\pi}}\left(  e^{-2\left(  x+2\right)  ^{2}/3}%
+e^{-2x^{2}/3}+e^{-2\left(  x-2\right)  ^{2}\backslash3}\right)  \text{,}%
\end{align*}
whose graph is Figure 8.
\begin{figure}
\centering
\includegraphics[keepaspectratio, width=5.00in]{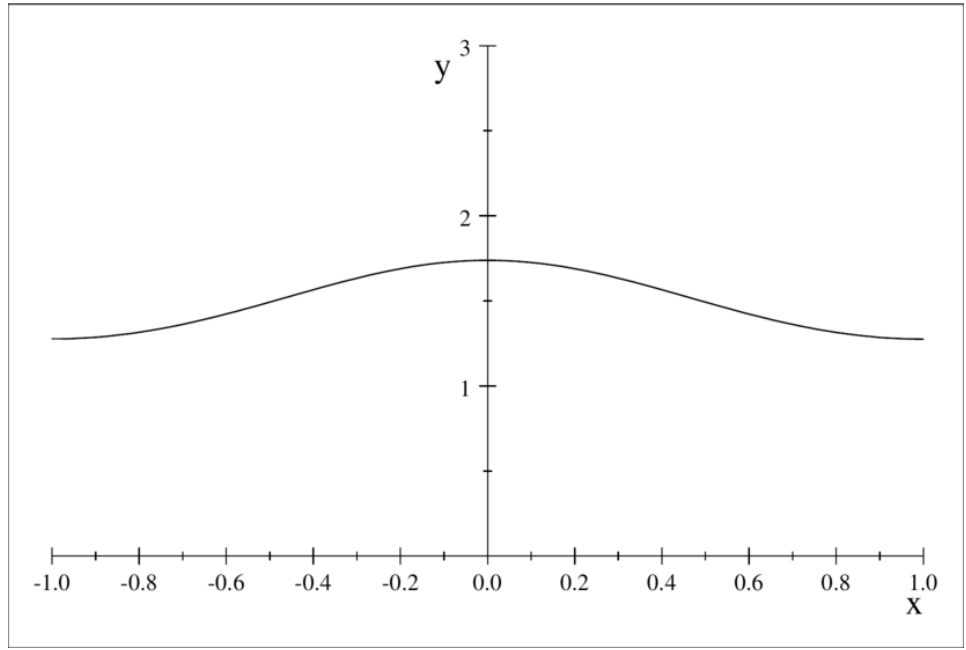}
\caption{$\sigma=1/2$, density after $3^{rd}$ primordium}
\end{figure}
And the minimum is still at $x=\pm1$. In fact the whole curve is much the same
as the previous curve, except that each point is higher (by about .$5$). This
is due to the fact that the distribution of antiauxin from the first
primordium has diffused to that of Fig. 5. The additional antiauxin will
presumably be absorbed. We see that repetitions of this process will give
essentially the same result, which is to say, the process is stable. This
represents distichous phyllotaxy. This stable distribution of antiauxin is not
flat as it was for $\sigma=1$, but flatter than it will be for smaller
$\sigma$, a possible explanation for the wraparound feature of distichous leaves.

\subsubsection{Spiral Phyllotaxy}

If $\sigma=1/4$, the 2$^{nd}$ primordium will again be at $\pm1$ but for the
3$^{rd}$ the situation is qualitatively different (Figure 9): There are two
minima at $\pm.506$%

\begin{figure}
\centering
\includegraphics[keepaspectratio, width=5.00in]{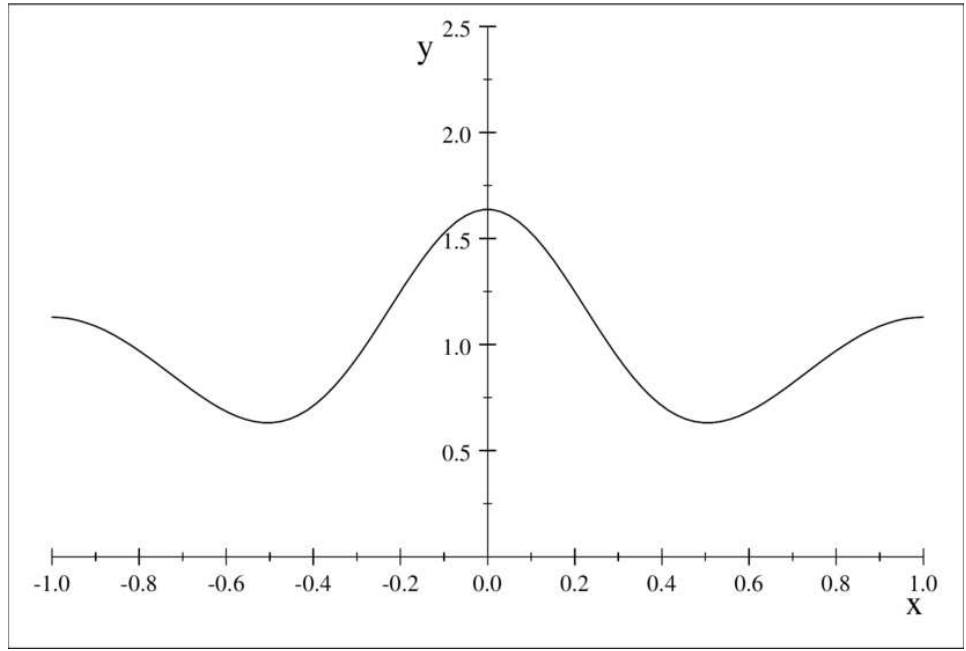}
\caption{$\sigma=1/4$, density after 2$^{nd}$ primordium}
\end{figure}
If we (arbitrarily) choose the positive minimum for the third primordium
(starting a right-handed spiral) and rotate it back to the origin, after
another plastochrone the density of antiauxin will be essentially%

\begin{align*}
&  \frac{4}{\sqrt{2\pi}}e^{-8x^{2}}+\frac{2}{\sqrt{\pi}}\left(  e^{-4\left(
x+.505\right)  ^{2}}+e^{-4\left(  x-1.495\right)  ^{2}}\right) \\
&  +\frac{4}{\sqrt{6\pi}}\left(  e^{-8\left(  x+1.505\right)  ^{2}%
/3}+e^{-8\left(  x-.495\right)  ^{2}/3}\right)
\end{align*}
whose graph is Figure 9.%
\begin{figure}
\centering
\includegraphics[keepaspectratio, width=5.00in]{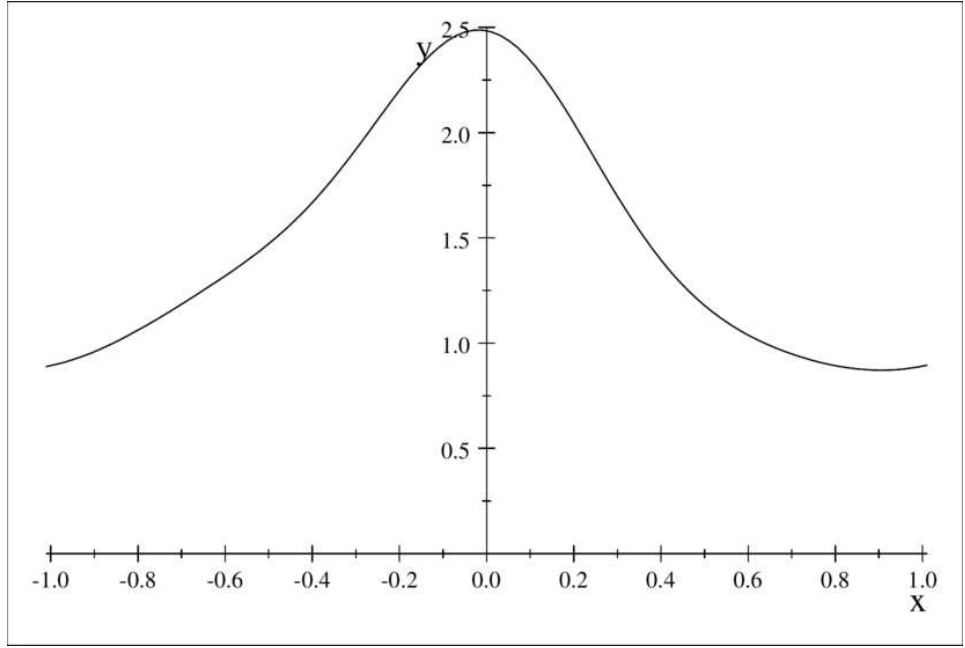}
\caption{$\sigma=1/4$, density after 3$^{rd}$ primordium}
\end{figure}

So, the (now unique) minimum is at $x_{3}=$ $.903$, continuing the
right-handed spiral. After another plastochrone the density is Figure 10.%
\begin{figure}
\centering
\includegraphics[keepaspectratio, width=5.00in]{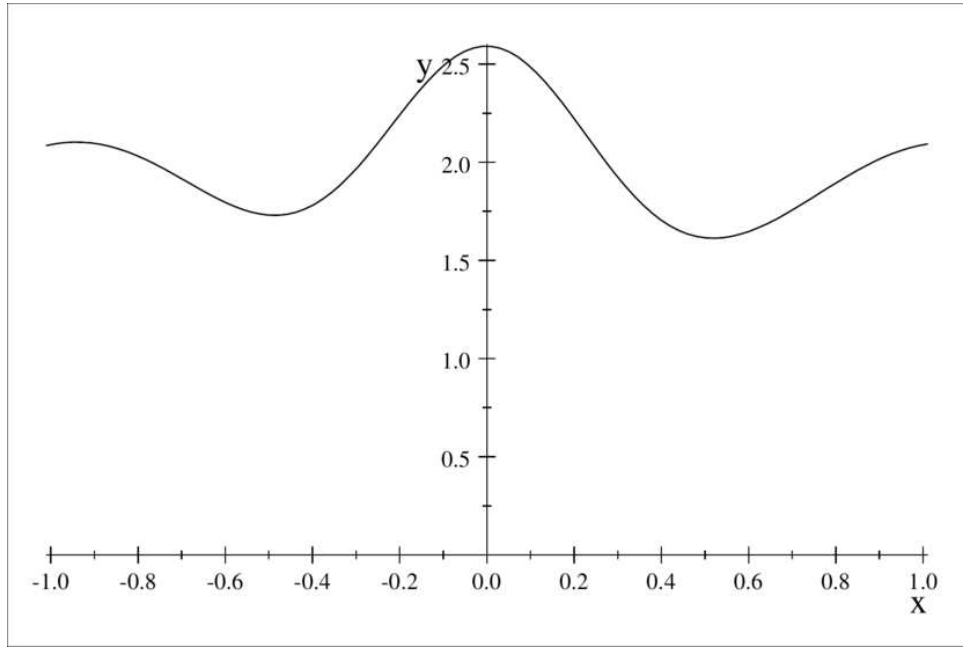}
\caption{$\sigma=1/4$, density after 4$^{th}$ primordium}
\end{figure}

So, the minimum is at $x_{4}=$ $.519$. The value $x_{k}$ is a measure of what
botanists call the divergence angle of the $k^{th}$ primordium (the angle that
it is turned from the previous primordium). Continuing on in this way we
obtained a sequence of divergence angles, the first 22 of which are: 1, .505,
.903, .519, .884, .534, .841, .546, .802, .560, .749, .577, .692, .596, .650,
.611, .630, .618, .625, .620, .623, .622. They then remain .622 (rounded off
to 3 decimal places to $n=30$ where we halted the computation. For all
practical purposes these values are convergent. This represents spiral
phyllotaxis. We shall return to the question of asymptotic convergence of the
divergence angles in Section 3.

The Smith \textit{et al} model of morphogenesis given in \cite{SGMRKP} is a
caricature of reality. Our model is a caricature of their caricature. It's
simplicity however, allows us to analyse it with calculus. It is not adequate
to account for the creation of the initial primordia. As was recognized by
Alan Turing \cite{Tur}, the initiation of primordia \textit{de novo} requires
a reaction that liberates energy. But does our simplified model retain some
essence of reality? Can it account for the placement of subsequent primordia
and their qualitative properties?

\subsection{Where Does Slow Diffusion Leave Off and Fast Diffusion Begin?}

We have seen in the previous examples that if $\sigma$ is large ($\geq1/2$)
the antiauxin density one plastochrone after the placement of the $k^{th}$
primordium will have only one minimum, right on top of the $(k-1)^{st}$
primordium. The $(k+1)^{st}$ primordium goes at that minimum and we have a
model of distichous phyllotaxy. On the other hand, if $\sigma$ is small
($\leq1/4$), one plastochrone after the $2^{nd}$ primordium the density
function has two symmetrical minima, each of which will (with probability
$1/2$) be the site of the 3$^{rd}$ primordium. This random choice initiates a
left- or right-handed spiral which subsequent placements continue. These
constitute the two kinds of mirror symmetric spiral phyllotaxy.

Evidently some $\sigma_{0}$, $1/4<\sigma_{0}<1/2$, is a critical point between
where the graph of auxin density at time 2 is concave down at $x=\pm1$ and
where it is concave up. Taking $x=+1$ to represent $x=\pm1$, this means (since
higher order wraparounds are negligeable) that $F(\sigma)=$
\begin{align*}
&  \frac{\partial^{2}(f(1;0,\sigma,1)+f(1;2,\sigma,1)+f(1;1,\sigma
,2)+f(1;-1,\sigma,2))+f(1;3,\sigma,2)}{\partial x^{2}}\\
&  <0\text{ for }\sigma<\sigma_{0}%
\end{align*}
and
\[
F(\sigma)>0\text{ for }\sigma>\sigma_{0}\text{.}%
\]
Thus we must solve $F(\sigma)=0.$ Now
\[
f(x;x_{0},\sigma,t)=\frac{1}{\sqrt{2\pi t}\sigma}e^{-\left(  x-x_{0}\right)
^{2}/2\sigma^{2}t}%
\]
so
\[
\frac{\partial f(x;x_{0},\sigma,t)}{\partial x}=\frac{1}{\sqrt{2\pi t}\sigma
}\frac{\partial e^{-\left(  x-x_{0}\right)  ^{2}/2\sigma^{2}t}}{\partial
x}=-\frac{1}{\sqrt{2\pi t^{3}}\sigma^{3}}\left(  x-x_{0}\right)  e^{-\left(
x-x_{0}\right)  ^{2}/2\sigma^{2}t}%
\]
and
\[
\frac{\partial^{2}f(x;x_{0},\sigma,t)}{\partial x^{2}}=\frac{1}{\sqrt{2\pi
t^{5}}\sigma^{5}}\left[  \left(  x-x_{0}\right)  ^{2}-\sigma^{2}t\right]
e^{-\left(  x-x_{0}\right)  ^{2}/2\sigma^{2}t}\text{.}%
\]
Therefore, since $f(x;x_{0},\sigma,t)$ is symmetric about $x_{0}$ and only
depends on $x-x_{0}$,%

\begin{align*}
F(\sigma)  &  =2\frac{\partial^{2}(f(1;0,\sigma,1)}{\partial x^{2}}%
+\frac{\partial^{2}(f(1;1,\sigma,2)}{\partial x^{2}}+2\frac{\partial
^{2}(f(1;3,\sigma,2)}{\partial x^{2}}\\
&  =\frac{1}{\sqrt{2\pi}\sigma^{5}}\left[  2\left(  1-\sigma^{2}\right)
e^{-1/2\sigma^{2}}+\frac{1}{2^{5/2}}\left(  \left(  -2\sigma^{2}\right)
+2\left(  2^{2}-2\sigma^{2}\right)  e^{-2^{2}/2\cdot2\sigma^{2}}\right)
\right] \\
&  =\frac{2}{\sqrt{2\pi}\sigma^{5}}\left[  \left(  1-\sigma^{2}\right)
e^{-1/2\sigma^{2}}+\frac{1}{2^{5/2}}\left(  -\sigma^{2}+\left(  2^{2}%
-2\sigma^{2}\right)  e^{-1/\sigma^{2}}\right)  \right] \\
&  =\frac{2^{-3/2}e^{-1/\sigma^{2}}}{\sqrt{2\pi}\sigma^{5}}\left[
2^{5/2}\left(  1-\sigma^{2}\right)  e^{1/2\sigma^{2}}-\sigma^{2}%
e^{1/\sigma^{2}}+2\left(  2-\sigma^{2}\right)  \right]
\end{align*}
So letting $s=\sigma^{2}$ we wish to solve
\[
4\sqrt{2}\left(  1-s\right)  e^{1/2s}-se^{1/s}+2\left(  2-s\right)  =0
\]%
\begin{figure}
\centering
\includegraphics[keepaspectratio, width=5.00in]{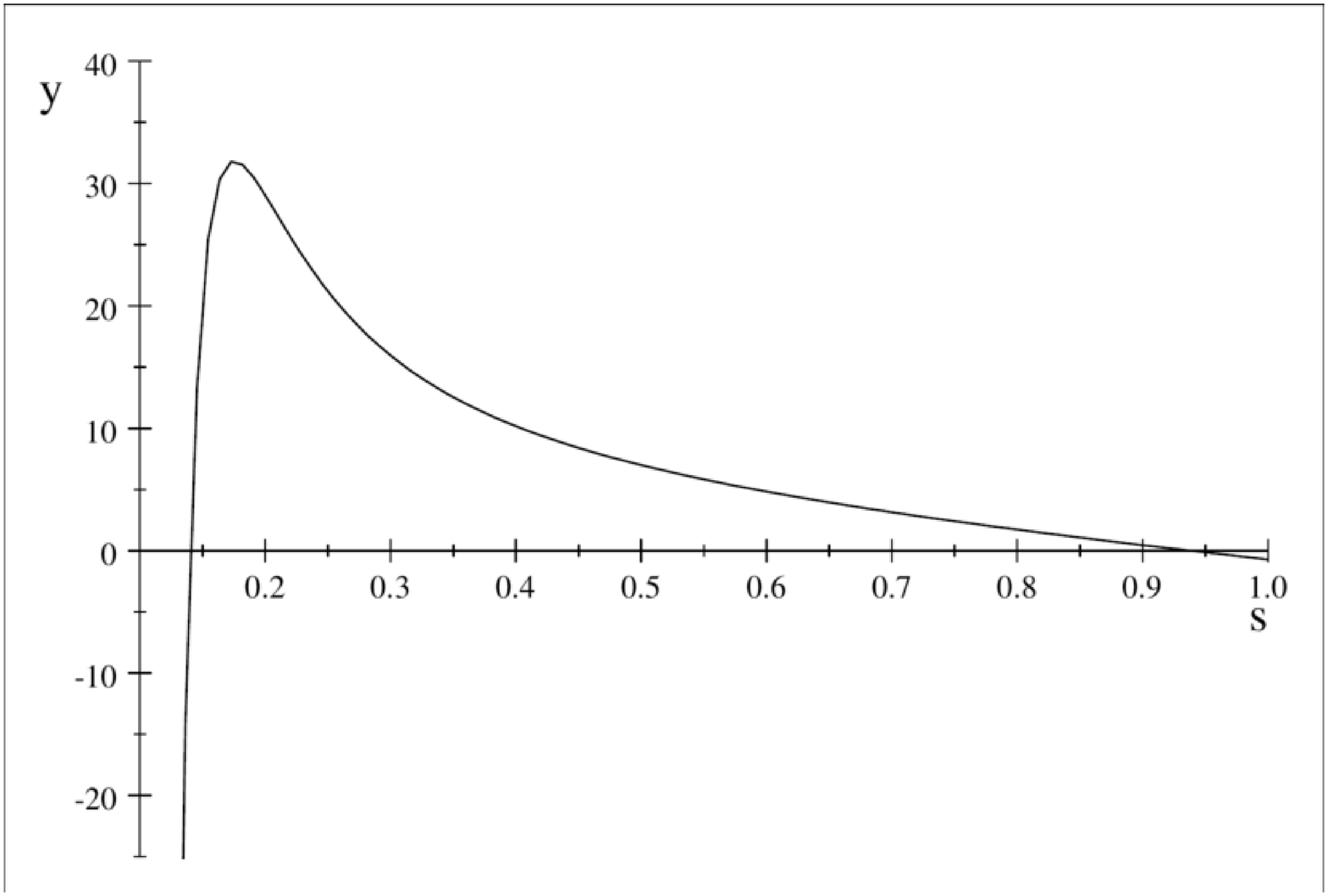}
\caption{The graph of $y=4\sqrt{2}\left(  1-s\right)  e^{1/2s}-se^{1/s}+2\left(  2-s\right)  $}
\end{figure}
and we see that $s_{0}$, the root we are looking for, is around $0.14$. Our
calculation showed that $s_{0}$ is exactly $0.140$ to 3 decimal places, so
$\sigma_{0}=\sqrt{.140}=\allowbreak0.374\,$ and $\frac{1}{3}<\sigma_{0}%
<\frac{1}{2}$ as we had surmised earlier. Note that the change of $F(\sigma)$
near $\sigma_{0}$ is precipitous.

\section{The Range of Divergence Angles}

\subsection{For Small $k$}

Here is a table of divergence angles from various sources for values of $k$,
the number of primordia on a stem:%

\begin{tabular}
[c]{r||rrrrrrrrrrrr}%
$k$ & $1$ & $2$ & $3$ & $4$ & $5$ & $6$ & $7$ & $8$ & $9$ & $10$ & ... &
$\infty$\\\hline
Arab. & $-$ & $170$ & $108$ & $158$ & $127$ & $145$ & $130$ & $135$ & $132$ &
$133$ & ... & $137.5$\\\hline
SGMRKP & $-$ & $160$ & $108$ & $157$ & $125$ & $143$ & $140$ & $130$ & $142$ &
$123$ & ... & \\\hline
$\sigma=1/4$ & $-$ & $180$ & $91$ & $162$ & $93$ & $159$ & $98$ & $152$ & $98$
& $144$ & ... & $112$\\\hline
$\ \sigma=1/3$ & $-$ & $180$ & $119$ & $146$ & $122$ & $141$ & $128$ & $136$ &
$130$ & $134$ & ... & $132$%
\end{tabular}

The first row, taken from the paper by Smith \textit{et al }\cite{SGMRKP},
shows divergence angles measured in \textit{Arabadopsis Thaliana, }the
standard subject for laboratory botanical experiments. The values are averages
with standard deviations from $8^{\circ}$ for $k=2$ and increasing up to
$20^{\circ}$. The second row, also from \cite{SGMRKP}, is the result of a
calculation with the SGMRKP (computer-based) model of spiral phyllotaxis. The
third and fourth rows are results of our calculations with the purely
diffusive model. $\sigma=1/4$ was the calculation we ran through in Section
2.4.2 but converted to degrees. Since the calculation of the previous section
showed that $\sigma=1/3<\sigma_{0}$ still gives spiral phyllotaxis, we
calculated the sequence of divergence angles for it also. Some observations:

\begin{enumerate}
\item The agreement between rows 1 (Arabadopsis) and 2 (the SGMRKP
computational model) is quite good for $3\leq k\leq6$, but then they drift apart.

\item Row 4 ($\sigma=1/3$) is not so close to row 1 (arabadopsis) at the
beginning but their agreement for $k\geq7$ is striking. Smith \textit{et al}
state that the divergence angle for Arabadopsis converges to the Fibonacci
Angle ($360/\tau^{2}\simeq137.5^{\circ}$where $\tau=\left(  1+\sqrt{5}\right)
/2$, the Golden Mean (See \cite{Jea94})). However row $4$ converges to
$132^{\circ}$ and the coconvergence of rows 1 \& 4 make it seem likely that
the limit of row 1 is more like 132.5$^{\circ}$. The claim that row 1
converges to $137.5^{\circ}$ may be a misuse of the mathematical theorem that
the Fibonacci angle "ensures an equitable allocation of space (for the organs
in spiral phyllotaxis)" \cite{Jea98}, p. 410. It is true that divergence
angles tend to cluster about the Fibonacci angle and and the classical
examples, the pine cone and sunflower capitulum, have divergence angles very
close to $137.5^{\circ}$. For many other examples however (such as the
Arabadopsis stem), where space is not scarce enough to apply selective
pressure, divergence angles vary widely.

\item In row 1 (Arabadopsis) the first angle should be about $180^{\circ}$
(assuming Davis's hypothesis is not valid and there is no environmental (or
intrinsic developmental) factor biasing the placement of the second primordium
to either side. In our calculations (rows 4 \& 5) there is no stochastic
component so this is the exact value. Smith \textit{et al }evidently took the
the divergence angle for the second primordium to be the length of the
\textbf{shortest} arc between the two. This is not unreasonable but it does
distort the underlying stochastic process a bit. In our pure model, since the
placement of the second primordium does not break mirror symmetry, the
handedness of the stem is determined by the third primordium. We arbitrarily
chose it to be on the right side, thus establishing a right-handed spiral. For
Arabadopsis the distribution of the first divergence angle is likely to be a
normal distribution with mean $180^{\circ}$ and standard deviation of about
$13^{\circ}$ (judging from the standard deviations given for the other
measurements). The standard deviation given for that first divergence angle in
row 1 ($8^{\circ}$) is considerably less than for all the other measurements.
If the angles had been measured in the positive sense (thereby allowing angles
greater than $180^{\circ}$), and they are normally distributed about
$180^{\circ}$, their mean deviation from $180^{\circ}$ would be about $\left(
\sqrt{\frac{\pi}{2}}\times\left(  180-170\right)  \right)  ^{\circ
}=\allowbreak12.\,\allowbreak533^{\circ}$. Also, their standard deviation
would be about$\frac{8}{\sqrt{1-\frac{2}{\pi}}}^{\circ}=\allowbreak
13.\,\allowbreak271^{\circ}$ (See Appendix for derivations of these formulas).
These inferences seem consistent with the measurements.

\item If row 1 had been started with $180^{\circ}$ (the likely average by
symmetry) instead of $170^{\circ}$, its agreement with row 4 would be even
more striking. We could also increase $\sigma$ a bit in row 4 to make the
limiting divergence angles coincide with $132.5^{\circ}$.

\item As pointed out in the previous remark (4), there is impressive agreement
between data from real plants and our simplified model in at least one case
(Arabadopsis \& $\sigma=1/3+$). The diffusion-only model captures the initial
oscillatory behavior of divergence angles as well as their eventual
convergence even better than the presumably more realistic model of Smith
\textit{et al}. Note that row 2 (SGMRKP) does not reliably oscillate nor
approach a limit.

\item In the beginning we assumed that the underlying diffusion process was
1-dimensional (confined to the periphery of the SAM). One might object to this
assumption since the tunica is 2-dimensional and the stem as a whole is
3-dimensional and auxin might well diffuse into those surrounding tissues. We
repeated the foregoing calculations with 2-\&3-dimensional diffusions
(replacing $t$ by $t^{2}$ \& $t^{3}$) and found that the results(oscillatory
behavior, limiting behavior) were qualitatively the same. The value of
$\sigma$ that produced a given limit divergence angle differed slightly from
dimension to dimension but at this point $\sigma$ has no physical reality
anyway. If at some future time $\sigma$ can be measured, then we may have to
modify the dimensionality of the diffusion. Another justification for the
1-dimensional model, besides its simplicity, is that it is the limit of the
2-dimensional model as internodal distances go to 0, and many plants do have
small (compared to the semicircumference) internodal distance (for definition
see \cite{Jea98} pg. 250 or Wikipedia).
\end{enumerate}

\subsection{Limit Divergence Angles as a Function of $\sigma$}

In the preceding table we listed limit divergence angles for $\sigma=1/4$ and
$1/3$. Calculating limit divergence angles for $.1\leq\sigma\leq.9$ in the
same way and plotting them we obtain the graph in Figure 13
\begin{figure}
\centering
\includegraphics[keepaspectratio, width=5.00in]{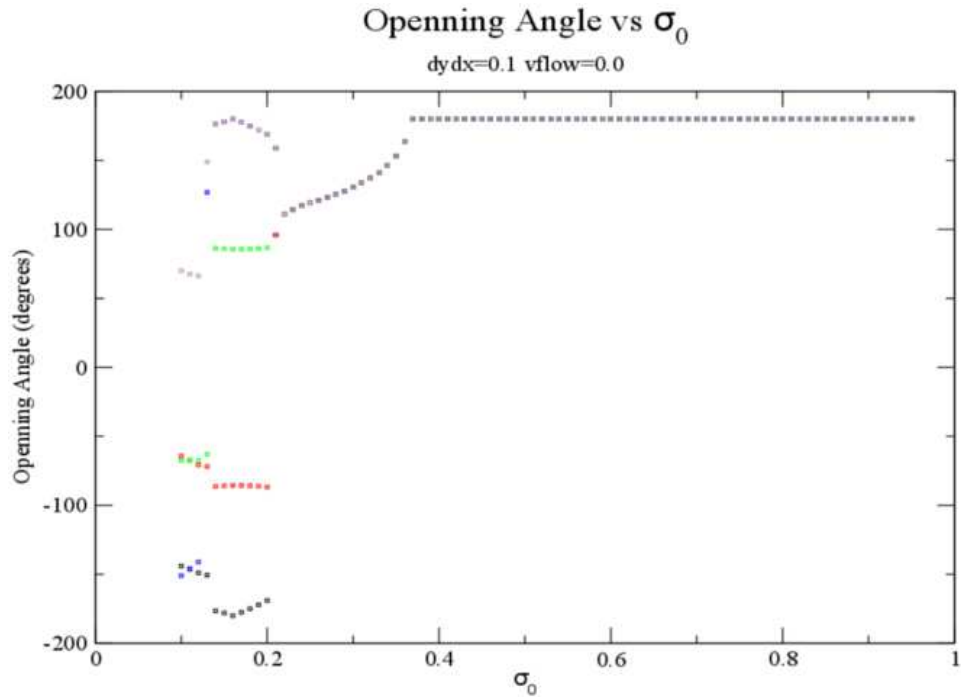}
\caption{Limit divergence angles}
\end{figure}
As expected, for $\sigma>\sigma_{0}=.374$ the limit divergence angles are all
$180^{\circ}$. For $\sigma<.25$ a kind of instability occurs, so there is no
limit for divergence angles but 4 different accumulation points. We have not
yet looked into this phenomenon but feel it is related to multijugate
phyllotaxis (next section). As mentioned earlier, the limiting divergence for
$\sigma=1/4$ is 112$^{\circ}$. Over the interval $.25\leq\sigma\leq.374$ the
limit divergence angle is smoothly increasing from 112$^{\circ}$ to
180$^{\circ}.$

In Chapter 27 of \cite{Jea98}, Meinhardt, Koch \& Bernasconi consider a
reaction-diffusion model of phyllotaxis. Their setup is similar to ours: The
basic ingredients being an activator, $a$ (auxin?) and two antagonists, $h$
and $s$, interacting in the peripheral zone of the SAM. In Chapter 18, Section
4 of \cite{Jea98}, Koch, Bernasconi \& Rothen elaborate on this model. Figure
9 of the latter article shows a plot of limit divergence angles as a function
of the diffusion constant of $a$. It looks similar to the plot above, even
though produced by a somewhat different model. However, there are also
quantitative differences: Their curve also splits into 4 branches below .25,
but the main one goes to the golden angle (137.5$^{\circ}$) as the diffusion
constant $a$ decreases to 0. Also, the critical value of the diffusion
constant (above which the diffusion angle is 180$^{\circ}$) is about 1
(compared to our $\sigma_{0}=.374$). Their scaling is different (the perimeter
of the peripheral region being 1 rather than 2) but this only makes the
difference greater. Anyway, these independent calculations reinforce what we
found in repeating our calculations many times with different parameters: The
qualitative properties of divergence angles are stable under a variety of
reasonable assumptions. Our simple model participates in this conscensus and
agrees qualitatively with observations of plants.

\section{Multijugate Phyllotaxis}

$n$-Jugate phyllotaxis starts with $n$ primordia, equally spaced around the
peripheral zone and continues in that way at each successive plastochrone.
Decussate is the 2-jugate analog of distichous: At each plastochrone two more
primordia appear, their axis at a right angle to that of the previous two. Can
there be 2-jugate or $n$-jugate spiral phyllotaxis? Most certainly, and they
can be represented with the same machinery we used for the unijugate case. The
only difference is that we specify the creation of $n$ primordia at each
plastochrone. $\sigma$ must then be divided by $n$ and the resulting
divergence angles will also be divided by $n$.%

\begin{figure}
\centering
\includegraphics[keepaspectratio, width=5.00in]{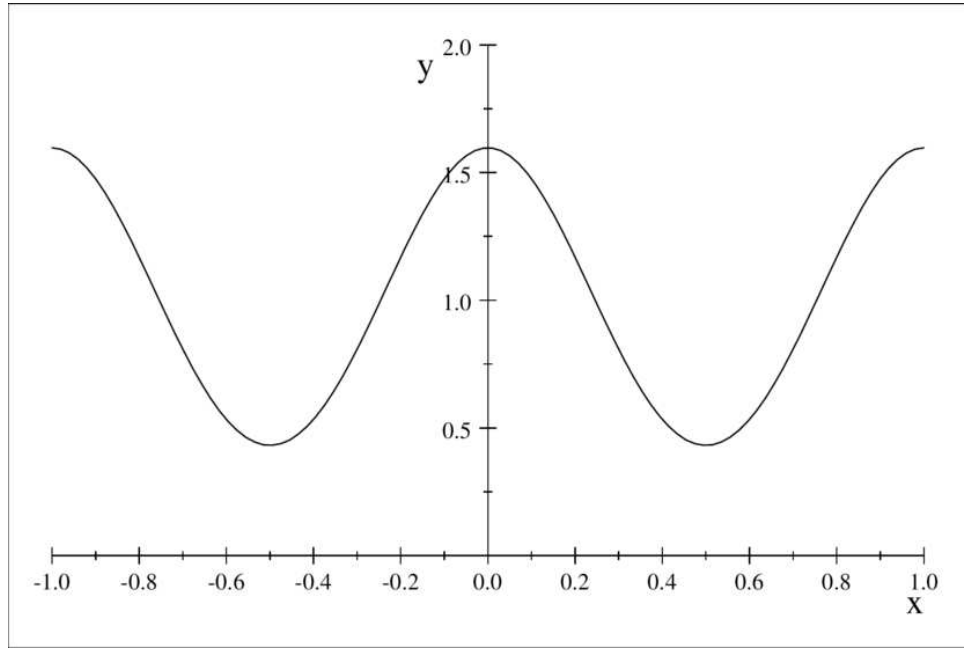}
\caption{A bijugate distribution for $\sigma=1/4$}
\end{figure}

Do the bases of decussate (or more generally n-cussate) leaves completely
encircle the stem? They do seem to in the few cases we have examined. However,
the bases of $n$-jugate \textbf{spiral} leaves cannot, if phyllotaxis is the
result of auxin dynamics, since that would block the flow of auxin. However,
as we showed in Section 2.4.1, it is not necessary for the bases of n-cussate
leaves to encircle their stem. It would be interesting to know if there are
any such and if not, why not?

\section{The Chirality of Branches}

Now we address the question that motivated this project: How is the asymmetry
of a stem passed on to a branch? Can auxin dynamics account for the fact that,
in a plant with spiral phyllotaxis, an axillary branch may have the same or
the opposite chirality as its parent stem? And that the correlation between
the two may be positive or negative? Figure 15 copies a graph from the paper
by Gomez-Campo \cite{G-C} showing the correlation between the handedness of a
branch at the axil of the $k^{th}$ prophyll and that of its parent stem in the
annual plant Hirschfeldia incana:

\begin{figure}
\includegraphics[keepaspectratio, width=5.00in]{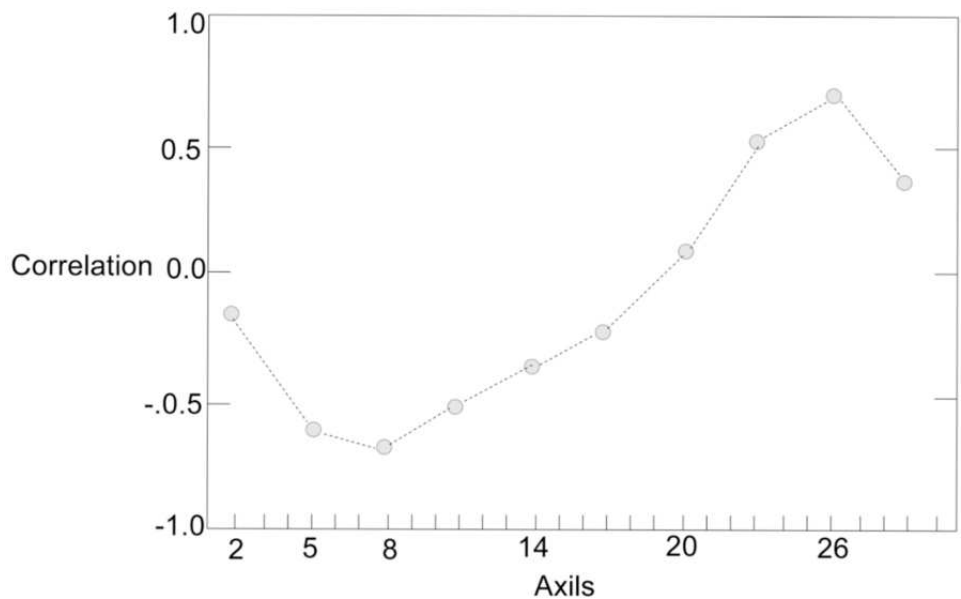}
\caption{Gomez-Campo's correlation data}
\end{figure}

We see a sinuous curve over the course of the growing season. It starts
slightly negative ($-0.2$) for the early branches, becomes more negative
reaching a minimum of $-0.7$ at $k=8$. It then increases, passing through
$0.0$ at $k=19$, reaching a maximum of $0.7$ at $k=26$ and dropping back to
$0.4$ at $k=29$. Previous authors had noted isolated correlations ranging from
$-1.0$ to $1.0$ and values in between, but Gomez-Campo was the first to record
a systematic change over time. At first this lability of corellation might
seem surprising since the handedness of stems is so unchanging with time
(exceptions have been observed, due to injury of the SAM or seasonal
dormancy). Gomez-Campo suggested that correlations could vary over time
because of the varying inhibitory influence of successor prophylls. Our model
gives quantitative substance to this idea.

First we must look more closely at the physiology of axillary branches. A
branch is physiologically the same as a stem. It has a SAM (we call it
SAM$_{k}$ if it is at the base of the $k^{th}$ leaf primordium (prophyll
P$_{k}$)), beginning with a few pluripotent cells that break off from the SAM
of the stem when P$_{k}$ separates from the peripheryl zone. They stick to the
upper edge of the prophyll but do not develop, probably because the prophyll
is using up all the auxin. At some instant (time $t_{k}$ for SAM$_{k}$)
sufficient auxin becomes available (maybe because the prophyll P$_{k}$
completes its growth or the SAM of the stem is excised) and the branch begins
to grow. We posit that the gradient of the antiauxin density at P$_{k} $ at
time $t_{k}$ influences the chirality (right or left) of the branch. This is
based on the following observations and assumptions:

\begin{enumerate}
\item The base diameter of the branch is relatively small compared to the stem
diameter at P$_{k}$. Therefore, any influence that the stem might have on
auxin concentrations in the branch will be mediated by the gradient of the
auxin concentration at P$_{k}$.

\item The incipient branch already has a prophyll, P$_{k,1}=$ P$_{k}$, on the
lower side of its base so its second prophyll, P$_{k,2}$ will be on the upper
side of the branch. For our model of a stem, the second prophyll is exactly
180$^{\circ}$ from the first so it is only the third prophyll whose placement
breaks the symmetry and determines whether the stem is right- or left-handed.
However, the placement of real stems have a stochastic component, so if the
placement of a second prophyll differs significantly from 180$^{\circ}$ it may
already determine the handedness of the stem. The same, of course, holds for
branches but there is another factor, the influence of other prophylls,
P$_{j}$, which may skew the distribution.

\item Assume that our stem has right-handed spiral phylotaxis. If the
horizontal gradient of the antiauxin density at P$_{k}$ is positive at the
moment when its axillary bud starts to grow, that will induce a positive
gradient across the base of the branch (as viewed from P$_{k,1}$). That
positive gradient will be promulgated out to the second primordium on the
branch and will tend to push P$_{k,2}$ to the left. If so the spiral will be
left-handed more than half the time, so stem and branch will be discordant.
The larger the gradient, the more likely the branch to be discordant. If, on
the other hand, the gradient is negative, the branch will tend to be
concordant. For left-handed stems the situation is the mirror image of that
for right-handed stems and probabilities of concordance and discordance are
the same.

\item In our model, when P$_{k}$ is established (at plastichrone $k$ by
definition) the derivative of the horizontal concentration of antiauxin is
$0.0$ since P$_{k}$ is placed at a minimum of antiauxin concentration. At that
time, P$_{k}=$ P$_{k,1}$ becomes a source of antiauxin which diffuses away
symmetrically. Therefore the derivative of antiauxin concentration will remain
$0.0$. As the antiauxin sources at P$_{j}$, $j<k$, age out, the antiauxin
concentration at P$_{k}$ may change a bit but it, and its derivative, will
remain close to $0.0$. This heuristic conclusion has been validated computationally.

\item However, after its initiation at plastochrone $k+1$, $P_{k+1},$ will
exert a nontrivial influence on the antiauxin concentration near $P_{k}$. If
the stem is right-handed, the gradient of the antiauxin concentration at
$P_{k}$ due to $P_{k+1}$ will be positive (since the divergence angle is
strictly between $0.0$ and $1.0$ ($0.0$ and 180$^{\circ}$).

\item The second successor, $P_{k+2}$, will generally have a negative
divergence angle (from $P_{k}$) of absolute value less than that of $P_{k+1}$.
So $P_{k+2}$, though further from $P_{k}$ vertically, will be closer
horizontally. The combined effect (the sum of the derivatives) may be positive
or negative depending on the geometry and the timing. That is, qualitatively,
the explanation for Gomez-Campo's observations.
\end{enumerate}

\subsection{The Quantitative Model}

We don't know much about Hirschfeldia Incana, so let us use the convenient
value $\sigma=1/3$, which gives a limit divergence of $132/180=\allowbreak
0.733\,$\ (from Section 3.1). The delay, $d$, will be the time (in
plastochrones) between the establishment of $P_{k}$ and $t_{k}$ (when $S_{k}$
starts growing). Implicit in our simplifying assumption of a fixed (limit)
divergence is that $k$ is fairly large. So what will be the gradient of the
antiauxin density at $P_{k}$ at time $t_{k}=k+d$ be? From our assumptions it
is only a function of $d$. For $0\leq d<1$ we have argued above (4) that it
will be essentially zero. At $d=1$ the influence of the antiauxin source at
$P_{k+1}$ will begin to be felt. For $1\leq d<2$ (assuming that the
contributions for all $P_{j}$, $j<k$, remain negligeable, the gradient of the
antiauxin density at $P_{k}$ will be%
\[
\frac{1}{\sqrt{2\pi}}\frac{.733}{\left(  1/3\right)  ^{3}\left(  d-1\right)
^{3/2}}\left(  e^{-\frac{\left(  .733\right)  ^{2}}{2\left(  1/3\right)
^{2}\left(  d-1\right)  }}\right)  -\frac{1}{\sqrt{2\pi}}\frac{2-.733}{\left(
1/3\right)  ^{3}\left(  d-1\right)  ^{3/2}}\left(  e^{-\frac{\left(
2-.733\right)  ^{2}}{2\left(  1/3\right)  ^{2}\left(  d-1\right)  }}\right)
\text{.}%
\]
The first term is the direct diffusion from $P_{k+1}$ to $P_{k}$, the second
is the wraparound. All other wraparounds are negligeable.

For $2\leq d<3$ we have the same formula for the contribution from $P_{k+1}$
plus the contribution from $P_{k+2}$. The total is%
\begin{align*}
&  \frac{1}{\sqrt{2\pi}}\frac{.733}{\left(  1/3\right)  ^{3}\left(
d-1\right)  ^{3/2}}\left(  e^{-\frac{\left(  .733\right)  ^{2}}{2\left(
1/3\right)  ^{2}\left(  d-1\right)  }}\right)  -\frac{1}{\sqrt{2\pi}}%
\frac{2-.733}{\left(  1/3\right)  ^{3}\left(  d-1\right)  ^{3/2}}\left(
e^{-\frac{\left(  2-.733\right)  ^{2}}{2\left(  1/3\right)  ^{2}\left(
d-1\right)  }}\right) \\
&  -\frac{1}{\sqrt{2\pi}}\frac{2-2\left(  .733\right)  }{\left(  1/3\right)
^{3}\left(  d-2\right)  ^{3/2}}\left(  e^{-\frac{\left(  2-2\left(
.733\right)  \right)  ^{2}}{2\left(  1/3\right)  ^{2}\left(  d-2\right)  }%
}\right)  +\frac{1}{\sqrt{2\pi}}\frac{2\left(  .733\right)  }{\left(
1/3\right)  ^{3}\left(  d-2\right)  ^{3/2}}\left(  e^{-\frac{\left(  2\left(
.733\right)  \right)  ^{2}}{2\left(  1/3\right)  ^{2}\left(  d-2\right)  }%
}\right)  \text{.}%
\end{align*}

Again, the first new term is the direct diffusion (this time from the left, so
it is negative) and the second the wraparound.

For $3\leq d<4$ we have the same formula for the contribution from $P_{k+1}$
and $P_{k+2}$ plus the contribution from $P_{k+3}$. The total is%
\begin{align*}
&  \frac{1}{\sqrt{2\pi}}\frac{.733}{\left(  1/3\right)  ^{3}\left(
d-1\right)  ^{3/2}}\left(  e^{-\frac{\left(  .733\right)  ^{2}}{2\left(
1/3\right)  ^{2}\left(  d-1\right)  }}\right)  -\frac{1}{\sqrt{2\pi}}%
\frac{2-.733}{\left(  1/3\right)  ^{3}\left(  d-1\right)  ^{3/2}}\left(
e^{-\frac{\left(  2-.733\right)  ^{2}}{2\left(  1/3\right)  ^{2}\left(
d-1\right)  }}\right) \\
&  -\frac{1}{\sqrt{2\pi}}\frac{2-2\left(  .733\right)  }{\left(  1/3\right)
^{3}\left(  d-2\right)  ^{3/2}}\left(  e^{-\frac{\left(  2-2\left(
.733\right)  \right)  ^{2}}{2\left(  1/3\right)  ^{2}\left(  d-2\right)  }%
}\right)  +\frac{1}{\sqrt{2\pi}}\frac{2\left(  .733\right)  }{\left(
1/3\right)  ^{3}\left(  d-2\right)  ^{3/2}}\left(  e^{-\frac{\left(  2\left(
.733\right)  \right)  ^{2}}{2\left(  1/3\right)  ^{2}\left(  d-2\right)  }%
}\right) \\
&  +\frac{1}{\sqrt{2\pi}}\frac{3\left(  .733\right)  -2}{\left(  1/3\right)
^{3}\left(  d-3\right)  ^{3/2}}\left(  e^{-\frac{\left(  3\left(  .733\right)
-2\right)  ^{2}}{2\left(  1/3\right)  ^{2}\left(  d-3\right)  }}\right)
+\frac{1}{\sqrt{2\pi}}\frac{3\left(  .733\right)  -4}{\left(  1/3\right)
^{3}\left(  d-3\right)  ^{3/2}}\left(  e^{-\frac{\left(  3\left(  .733\right)
-4\right)  ^{2}}{2\left(  1/3\right)  ^{2}\left(  d-3\right)  }}\right)
\end{align*}
Note that the formulae have several points ($d=1,2,3,4$) at which division by
$0$ takes place. However, the curve is smooth and the values at those points
may be calculated by interpolation.Thus the graph of the gradient of antiauxin
density for $1\leq d\leq4$ is Figure 16.%
\begin{figure}
\centering
\includegraphics[keepaspectratio, width=5.00in]{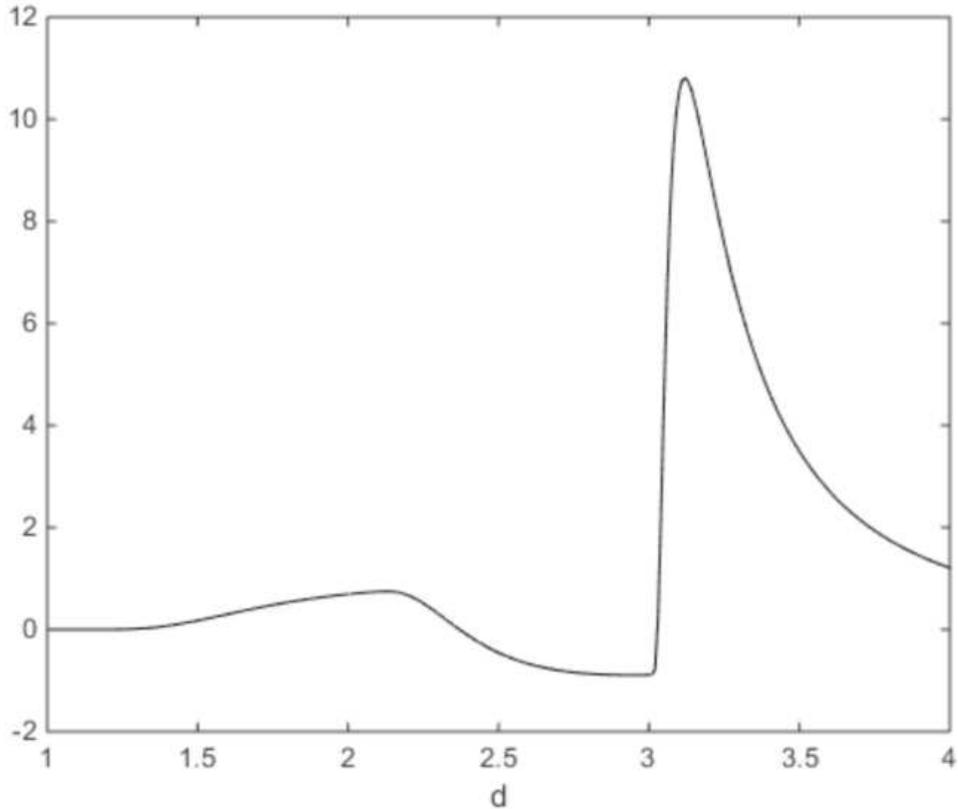}
\caption{The gradient of antiauxin density, $1\leq d\leq4$}
\end{figure}
So how does this connect with Gomez-Campo's correlation data (Figure 15)?
Recalling that positive antiauxin gradient leads to negative correlation of
handedness, we reflect Figure 16 about the $x$-axis (multiply $y$-values by
$-1$). We could call the resulting $y$-values "auxin gradient". Because of the
relatively close proximity of the $3^{rd}$ primordium (the angle is
$36^{\circ}$ compared to $132^{\circ}$ for the $1^{st}$ and $96^{\circ}$ for
the second), the contribution of the $3^{rd}$ primordium (after a short time)
is relatively large. At this point all we can say for sure about the
relationship between auxin gradient and correlation is that $0$ gradient
should give $0$ correlation and the larger the gradient, the larger the
correlation. However the gradient can take values from $-\infty$ to $\infty$
and correlation is restricted to between $-1$ to $1$. So trimming down the
domain of Figure 16 to $1\leq d\leq3.2$ and the range from $-1$ to $1$ we have
Figure 17.
\begin{figure}
\centering
\includegraphics[keepaspectratio, width=5.00in]{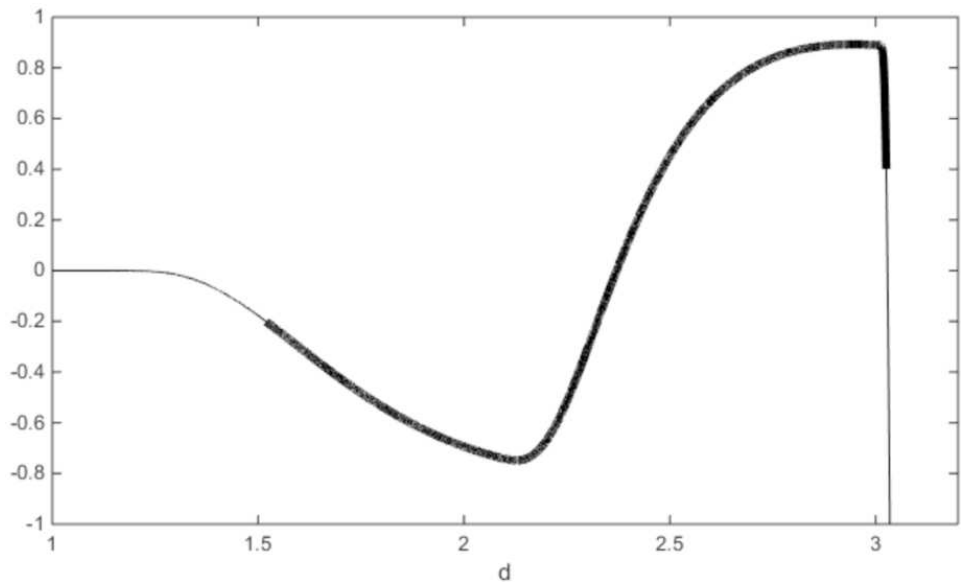}
\caption{Auxin gradient at a primordium after $d$ plastichrones, $1\leq d\leq3.2$}
\end{figure}

The heavy part of the curve was selected so that it goes from the same
$y$-value, $-0.2$, that Gomez-Campo's starts with, to the same $y$-value,
$0.4$, that Gomez-Campo's ends with. The similarity between the two curves
(Figure 15 and the dark part of Figure 17) is, we claim, evidence that auxin
dynamics in a stem heavily influences the relative handedness of a branch. The
biggest difference is the sharper drop after $d=3$ in Figure 17. This is
clearly an artifact of our oversimplified model (its 1-dimensionality). Adding
in another dimension (or two) would almost certainly smooth out that sharp
downturn. Also, increasing $\sigma$ in the model would increase the angular
distance to the $3^{rd}$ primordium and soften that hard turn.

If low values (near $0$, but positive and negative) can influence the
handedness of a branch, then high values (as seen for $d>3$ in Figure 16)
should essentially determine it). There are plants exhibiting each of the
extremes (always concordant or always discordant). How those plants differ
physiologically is an interesting question. Another puzzle is why these
differences (always concordant or always discordant or varying from discordant
to concordant) should be built into the development process? The classical
"explanation" of spiral phyllotaxis and the divergence angle of $137.5^{\circ
}$ is that it optimizes leaves' exposure to sunlight. As the measurements of
Arabadopsis thaliana show, some (probably most?) plants do not achieve that
ideal divergence angle. Distichous plants are not even close. So are these
exceptions subject to other selective pressures which causes them deviate from
the ideal? We have noticed that some distichous plants (Bananas, Bird of
Paradise) grow in clumps and that the outer stems orient their vertical plane
tangentially to the circular mass of stems. This would seem to maximize the
exposure of their leaves to the sun. Could there be similar causes for choices
of relative chirality?

Obviously, our model of the Gomez-Campo phenomenon is a shot in the dark, but
hopefully the beginning of a productive discussion.

\section{Conclusions and Comments}

\subsection{Exceptions?}

We believe that our model of morphogenesis based on diffusion-driven auxin
implies Hofmeister's Rule. Though infrequent, there are exceptions to
Hofmeister's Rule in the botanical literature (See \cite{R-P}, p.102). One
such is monostichy where all the leaves on a stem are aligned on one side.
Corn, with its rows of kernels, might seem to be an exception but closer
examination of corn shows that the kernels of one row are offset by half a
kernel from the neighboring row. 8-row corn (Golden Bantam) is actually
4-jugate and distichous. If corn is n-jugate and distichous then the number of
rows on a cob is even, which is generally true as the authoratative website
\[
\text{http://www.agry.purdue.edu/ext/corn/news/timeless/earsize.html}%
\]
attests. Our guess is that closer inspection of the other apparent exceptions
will reveal similar explanations. For instance monostichous plants may have
been initiated as distichous but then the primordia on one side not allowed to
develop. If this is so, there should be some traces of those primordia that
don't develop ("Ontogeny recapitulates phylogeny"-Ernst Haeckel).

We have had similar experiences several times already. In setting out to
examine the implications of diffusion-driven auxin, we initially thought that
it would preclude distichous phyllotaxis (as mentioned in Section 2.3) and
found to our surprise that distichous phyllotaxy fit right into the model. We
had not considered the possibility of random phyllotaxis because it also
seemed to violate Hofmeister's Rule. But since the model allowed it we looked
in the literature and there it was: \cite{R-P}, p.102. Most trees, pines,
oaks, figs, \textit{etc., }seem to have random phyllotaxy. In the beginning of
our fascination with phyllotaxy we ignored them, saying, "Plants are like
people. Some are better at mathematics than others!" It was a revelation to
find that their random phyllotaxy could follow from the same principle as
Fibonacci spirals. And the plants with random phyllotaxy may have an important
evolutionary advantage. Spiral phyllotaxy can be traced back at least to
ancient seas with brown algae. Grasses (with distichous phyllotaxy) were the
first plants to grow on land. Trees are a relatively recent development. They
evidently compete in crowded environments by randomly sending out branches and
then withdrawing nourishment from those that are not productive (of energy
from the sun). The "better mathematicians" don't seem to have that same
ability to adapt to their circumstances.

Green, Steele \& Rennich (\cite{Jea98}, Chapter 15) considered mechanical
buckling of the tunica as an alternative to chemical dynamics (such as our
diffusion driven auxin) in explaining phyllotactic patterns. Among the
successes they claim is the representation of organs "in line" (\cite{Jea98},
p. 381, Fig. 6A-B). They go on to say, "Interestingly, the natural, 'in line'
production of organs, seen rarely in flower development (Sattler, 1973;
Lacroix and Sattler, 1988), has been presented as special challenge to
modelers of phyllotaxis". It is a special challenge exactly because it
contradicts Hofmeister's rule and is not possible with our model of auxin
dynamics. However, flowers have clearly been subjected to more selection
pressure than other parts of plants, so one would expect their development to
be more complex. This needs further study.

\subsection{Distichous to Spiral?}

In Section 1.2.1 we modeled distichous phyllotaxis with $\sigma=1/2$ and later
argued that any value of $\sigma>\sigma_{0}=\allowbreak0.374$ would produce
distichous phyllotaxis and any value of $\sigma<\sigma_{0}$ would produce
spiral phyllotaxis. However, when we ran the calculation for $\sigma=.35$ it
was distichous for the first 15 primordia and then transitioned into spiral
over the next 5 primordia with a limiting divergence of 153$^{\circ}$. Clearly
the transition from distichous to spiral is more complex than represented in
Section 1.2.1. We think what happened at $\sigma=.35$ was that the small bump
visible in the graph of Figure 5, though inconseqential by itself, added up
over the generations of primordia to something not inconsequential. The closer
one gets to the transitional value $\sigma_{0}=\allowbreak0.374$, the longer
the transition from distichous (divergence angle of 180$^{\circ}$) to spiral
takes. For $\sigma=.36$ it only begins at the 50$^{th}$ primordium and
requires about 15 more plastochrones to complete. We have observed that
gladiolas exhibit a slow transition from distichous to spiral, ending up with
large divergence angles. It could be interesting to study such plants to see
if there is a consistent connection between their limit divergence angles and
speed of convergence to it.

\subsection{Does Nature Favor Concordance or Discordance?}

A casual examination of the data gathered by Gomez-Campo for
\textit{Hirschfeldia incana} shows that over the year more branches are
discordant than concordant. The average corellation is $-0.055$. The average
of all the average corellations for the 6 plants studied by Gomez-Campo is
$-.105$. This suggests that the delay, $d$, between the initiation of a
primordium and the initiation of its axillary branch for these plants is
mostly between $1.5$ and $2.25$ plastichrones, when the influence of the
immediate successor dominates. This tendency toward discordance was a surprise
since the stability of handedness in a stem is so universal and our
examination of bottlebrush had shown a corellation of about $1/3$.

\subsection{Possibilities}

We believe we have found a window through which the inner workings of plant
stems may be viewed. The possibilities for refinement and variation of our
model of the Gomez-Campo data are manifest and manifold. Here are a few:

\begin{enumerate}
\item Determine the actual limit divergence angle for Hirschfeldia incana
(equivalent to determining the corresponding value of $\sigma$. See Figure
13). Then redo the calculation of Section 5.1 with those values. Hopefully,
the resulting curve will be a better fit for that of Figure 15.

\item If we begin with small values of $k$ (the ordinal of the $k^{th}$
primordia, $P_{k}$) the alternation of initial divergence angles that showed
up in our calculations of Section 2.4.2 and in the table of Section 3.1 should
also effect the derivative of the gradient as a function of delay, $d$.
Gomez-Campo's corellation data for Hirschfeldia incana (Figure 15) only has
values for $k=2,5,8,11,14,17,20,23,26 \& 29$ so those alternations, if they
exist, have been averaged out. More and better data is called for.

\item The Gomez-Campo data was soley from plants for which branching was a
natural part of development. Cutting off the SAM of a stem can force branches
(called adventitious branches) to grow, even on stems that normally would not
branch. Presumably this happens because a major sink of auxin has been
removed. Does the timing and placement of such surgery effect the corellation
of chiralities? Recently we examined some stem-branch pairs in \textit{xylosma
senticosa, }a decorative shrub on the UCR campus which is heavily trimmed. Out
of a dozen adventitious branches, all were discordant. Can our model shed any
light on that?

\item As we noted in Section 3.1(5), the SGRMKP (computer based) model of
phyllotaxis does not seem to capture the alternation of divergence angles or
their convergence as $k$ becomes large. Is it possible to hybridize SGRMKP
model with ours and improve both?

\item The data from measurements on real plants has a stochastic component,
probably inherited from the underlying processes such as diffusion. However,
our model, based on the mathematics of diffusion, is deterministic. Could
incorporation of stochastic processes into our model make it more accurate?
For instance, rather than placing the next primordium at the minimum of the
antiauxin density, we might distribute it according to a distribution whose
density is a monotone function of the antiauxin density. So the lower the
antiauxin density, the greater the probability of the primordium being
initiated at that point. But what distribution would make the output a better
match for reality? Also, it seems reasonable that the distribution should
arise from the process for initiating a primordium.
\end{enumerate}

\subsection{In Summation}

The thing that seems most impressive about our oversimplified model is that it
accounts for so many mysterious characteristics of plant stem morphogenesis
with just one parameter, $\sigma$. In the model, variations of that one
parameter account for the magnitude of divergence angles, their oscillations
and limiting values and may well account for different corellations between
the handedness of a stem and its branches. $\sigma$ itself is not so readily
observable but the other parameters are. If they really do depend on $\sigma$,
there should be nonobvious relationships between them which can be verified
(or falsified) by direct measurements. If the model passes such tests, then
lab experiments to directly verify the role of auxin density could be
undertaken, opening up the prospect of guiding plant stem morphogenesis.

\section{Appendix}

If $X$ is a normal (Gaussian) random variable of mean $0$ and standard
deviation $\sigma$, then the mean of the absolute value of $X$ is
\begin{align*}
E\left(  \left\vert X\right\vert \right)   &  =\frac{1}{\sqrt{2\pi}\sigma}%
{\displaystyle\int\limits_{-\infty}^{\infty}}
\left\vert x\right\vert e^{-x^{2}/2\sigma^{2}}dx\\
&  =\frac{1}{\sqrt{2\pi}\sigma}2%
{\displaystyle\int\limits_{0}^{\infty}}
xe^{-x^{2}/2\sigma^{2}}dx\\
&  =\sqrt{\frac{2}{\pi}}%
{\displaystyle\int\limits_{0}^{\infty}}
ue^{-u^{2}/2}\sigma du\text{ if }x=\sigma u\\
&  =\sqrt{\frac{2}{\pi}}\sigma%
{\displaystyle\int\limits_{0}^{\infty}}
ue^{-u^{2}/2}du\\
&  =\sqrt{\frac{2}{\pi}}\sigma%
{\displaystyle\int\limits_{0}^{\infty}}
e^{-z}dz\text{ if }u^{2}/2=z\\
&  =\sqrt{\frac{2}{\pi}}\sigma\left[  -e^{-z}|_{0}^{\infty}\right] \\
&  =\sqrt{\frac{2}{\pi}}\sigma\left[  -\left(  0-1\right)  \right] \\
&  =\sqrt{\frac{2}{\pi}}\sigma\text{.}%
\end{align*}
So
\[
\sigma=\sqrt{\frac{\pi}{2}}E\left(  \left\vert X\right\vert \right)  .
\]

Also, the variance of the absolute value of $X$ is
\begin{align*}
Var\left(  \left\vert X\right\vert \right)   &  =E\left(  \left(  \left\vert
X\right\vert -E\left(  \left\vert X\right\vert \right)  \right)  ^{2}\right)
\\
&  =E\left(  \left\vert X\right\vert ^{2}\right)  -E^{2}\left(  \left\vert
X\right\vert \right) \\
&  =E\left(  X^{2}\right)  -E^{2}\left(  \left\vert X\right\vert \right) \\
&  =\sigma^{2}-\left(  \sqrt{\frac{2}{\pi}}\sigma\right)  ^{2}\\
&  \left(  1-\frac{2}{\pi}\right)  \sigma^{2}\text{.}%
\end{align*}
So%
\[
\sigma=\frac{\sqrt{Var\left(  \left\vert X\right\vert \right)  }}%
{\sqrt{1-\frac{2}{\pi}}}\text{.}%
\]

\end{document}